\def\cm{{\rm\thinspace cm}}

\def\erg{{\rm\thinspace erg}}

\def\keV{{\rm\thinspace keV}}

\def\Msun{\hbox{$\rm\thinspace M_{\odot}$}}

\def\s{{\rm\thinspace s}}

\def\ergcmps{\hbox{$\erg\cm\ps\,$}}

\def\ps{\hbox{$\s^{-1}\,$}}

\documentclass[usegraphicx]{mn2e}

\usepackage{fixltx2e}
\usepackage{mathptmx}

\include{defn}
\begin{document}

\title[Reflection and spin in Cyg X-1]{On the determination of the
  spin of the black hole in Cyg X-1 from X-ray reflection spectra}
\author[A.C. Fabian et al] {\parbox[]{6.5in}{{
      A.C. Fabian$^1\thanks{E-mail: acf@ast.cam.ac.uk}$,
      D. Wilkins$^1$, J.M.~Miller$^2$, R.C. Reis$^2$,
      C.S.~Reynolds$^3$, E.M.~Cackett$^{1,4}$, M.A.~Nowak$^5$,
      G.~Pooley$^6$, K.~Pottschmidt$^{7,8}$, J.S.~Sanders$^1$,
      R.R.~Ross$^9$ and J.~Wilms$^{10}$
        }\\
    \footnotesize
    $^1$ Institute of Astronomy, Madingley Road, Cambridge CB3 0HA\\
$^2$ Dept. of Astronomy, University of Michigan, ann Arbor, MI
48109, USA\\
$^3$ Dept. of Astronomy, University of Maryland, College Park,
MD 20742, USA\\
$^4$ Dept. of Physics and Astronomy, Wayne State University, Detroit,
MI 48201, USA\\
$^5$ Kavli Institute for Astrophysics and Space Research, MIT, 77
Massachusetts Avenue, Cambridge, MA02139, USA\\
$^6$ Cavendish Laboratory, University of Cambridge, JJ Thomson Avenue,
Cambridge CB3 0HE\\
$^7$ CRESST and NASA Goddard Space Flight Center, Code 661, Greenbelt
MD 20771, USA\\
$^8$ Center for Space Sciemce and Technology, University if Maryland
Baltimore County, 1000 Hilltop Circle, Baltimore MD21250, USA\\
$^9$ Physics Dept, College of the Holy Cross, Worcester, MA 01610,
USA\\
$^{10}$ Dr. Karl-Remeis-Sternwarte and Erlangen Center for
Astroparticle Physics, Sternwartstr. 7, 96049 Bamberg, Germany\\ 
 }}

\maketitle
  
\begin{abstract}
  The spin of Cygnus X-1 is measured by fitting reflection models to
  Suzaku data covering the energy band 0.9--400~keV. The inner radius
  of the accretion disc is found to lie within 2 gravitational radii
  ($r_{\rm g}=GM/c^2$) and a value for the dimensionless black hole
  spin is obtained of $0.97^{+0.014}_{-0.02}$. This agrees with recent
  measurements using the continuum fitting method by Gou et al. and of
  the broad iron line by Duro et al.  The disc inclination is measured
  at $23.7^{+6.7}_{-5.4}$~deg, which is consistent with the recent
  optical measurement of the binary system inclination by Orosz et al
  of $27\pm0.8$~deg. We pay special attention to the emissivity
  profile caused by irradiation of the inner disc by the hard
  power-law source. The X-ray observations and simulations show that
  the index $q$ of that profile deviates from the commonly used,
  Newtonian, value of 3 within $3r_{\rm g}$, steepening considerably
  within $2r_{\rm g}$, as expected in the strong gravity regime.
  
\end{abstract}


\section{Introduction}

Astrophysical black holes are characterised by just mass and spin. The
measurement of spin requires observations which probe the immediate
environment of the black hole event horizon. Two methods which have
emerged using current X-ray observations of luminous accreting black
holes rely on fitting either the shape of the reflection component of
the spectrum including a broad iron line, or the shape and the flux of
the quasi-blackbody continuum. Both assume that the accretion disc
extends in to the Innermost Stable Circular Orbit (ISCO), the radius
of which is determined by the black hole spin. The Reflection Fitting
method essentially measures the largest gravitational redshift of the
disc, which comes from the ISCO, and yields that radius in units of
gravitational radii ($r_{\rm g}=GM/c^2$). The Continuum Fitting method
measures the area of the inner disc and obtains the radius of the ISCO
in km. Here we apply the reflection method to Suzaku X-ray data of the
first stellar mass black hole, Cyg X-1.

Early attempts to measure the spin parameter of Cyg X-1 using the
reflection method led to inconsistent results (Miller at al 2005;
Miller et al 2009). Adopting a Newtonian emissivity profile (surface
flux on the disc varies with radius $r$ as $r^{-q}$ with $q=3$) and
using a special timing mode of XMM, a recent measurement by Duro et
al. (2011) gives a dimensionless spin parameter\footnote{The
  dimensionless spin parameter $a=cJ/GM^2$, where $J$ and $M$ are the
  angular momentum and mass of the black hole, respectively.} of
$a=0.88 ^{+0.07}_{-0.11}$.  Recent results using the Continuum Fitting
method have been obtained by Gou et al (2011). They find a near
extreme black hole with $a>0.95.$ The work relies on accurate optical
determinations of the mass ($14.8\pm0.1\Msun$) and inclination
($27\pm0.8$~deg) obtained by Orosz et al (2011).

Here we study the reflection/iron line approach in detail, taking
account of the relativistic effects on the emissivity profile expected
if the black hole spins rapidly. Such effects demand that the
emissivity be steep if the disc extends to the innermost region within
$\sim 2r_{\rm g}$ (Wilkins \& Fabian 2011). We find
$a=0.97^{+0.014}_{-0.02}$, which is a higher than, but marginally
consistent with, that reported by Duro et al (2011). Both reflection
results are consistent with the Continuum Fitting result of Gou et al
(2011).
  
We find the disc to be fairly highly ionized with an ionization
parameter around 1800~erg~cm~s$^{-1}$, which means that the reflection
spectrum is dominated by a large edge due to ionized iron. This
blurred edge carries information about spin in the spectral fit. Early
work emphasised how this edge can be important in fitting the spectrum
of Cyg X-1 (Ross, Fabian, Brandt 1996).

\section {Spectral fits}

The dataset we use is one of the 20 Suzaku observations of Cyg X-1
analyzed by Miller et al (2012; the observation of 2009 Apr 8). The
focus of that work is not on measuring the black hole spin, but rather
studying the detailed disc-jet coupling in Cyg X-1.  The spectra in
that work have been fitted with a relativistically-blurred reflection
model, but in order to make the work tractable, the fits kept the
source inclination and iron abundance frozen at 27 deg and 1,
respectively. The emissivity profile of the blurring function was also
just a single power-law, fixed for the Newtonian value of 3.  We relax
these assumptions here and apply them to an average dataset. We later
tested the model on several  other datasets (2009 May 6, 25 and
Dec 1) and obtained consistent results.

The best fit to the 1--500~keV spectrum is shown in Fig.~1 and values
of key parameters are given in Table~1. The model used, {\sc
  constant*tbabs*mtable(windabs.fits)(gaussian+gaussian
  +diskpbb+powerlaw+kdblur2f*atable(extendx.mod)) highecut}, is shown
in Fig.~2. {\sc kdblur2f} is a fast adaptation by JSS of the broken
power-law emissivity convolver {\sc kdblur2}. {\sc Extendx} is a
version of the self-consistent slab reflection model {\sc reflionx}
(Ross \& Fabian 2005) extended to allow the photon index to be less
than 1.4. (We do not use the {\sc refhiden} models of Ross \& Fabian
2007, which incorporate a blackbody component because they are only
available for Solar abundance iron and at the blackbody temperature
found here for the low state of about 0.16~keV the differences are
minimal.)  Absorption features due to the stellar wind of the
companion (e.g., Hanke et al. 2009, Nowak et al., 2011) are modeled
using {\sc windabs.fits} (Miller et al., 2012) and we refer to that
paper for further details. The absorption has little effect on our
results.  We note that the {\sc highecut} model used to fit the data
above 100~keV is phenomenological and is not strictly consistent with the
300~keV exponential cutoff assumed by the {\sc extendx} model, but has
limited effect on the results presented here. The quality of the fit
is very good, and the emissivity index $q$ is found to be high
$>6.8$. The inclination of $23.7^{+6.7}_{-5.4}$~deg is consistent with
the optical measurement of Orosz et al (2011).

It is clear that the spectrum is reflection-dominated below 10~keV
(Fig.~2), although the whole source, from an energy point of view
integrating over the whole spectrum, is not reflection dominated. This
is a consequence of the hard irradiating continuum where much of the
energy absorbed from Compton recoil due to photons at $\sim 100\keV$
is emitted by the surface layers if the disc below 6~keV.  The ratio
of the 0.1--1000~keV flux from the reflection component is 0.70 times
that from the power-law component, so the reflection fraction,
calculated as the ratio of incident to emitted flux at the disc
surface, ${\cal R}=0.7$. The blackbody disc component is 0.4 times the
power-law and if reflection is responsible for some of that emission
then ${\cal R}\sim 1$.  A narrow emission line is introduced at
6.4~keV to account for (weak) distant reflection (e.g. from the
stellar companion) and a narrow absorption feature is included at
1.85~keV to compensate for problems in the model response near the
detector silicon edge.
  
\begin{figure}
  \centering
  \includegraphics[width=0.7\columnwidth,angle=-90]{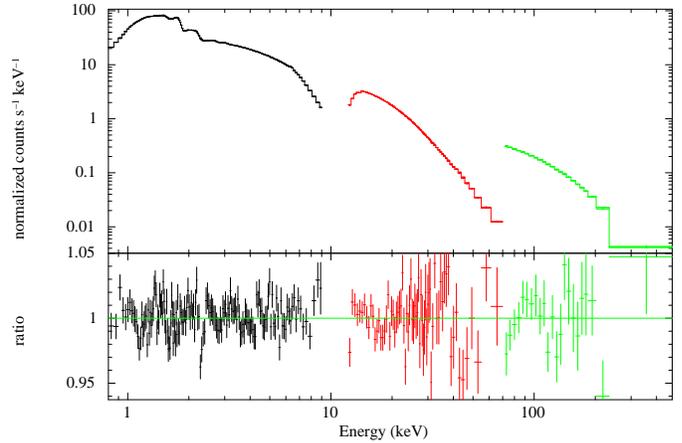}
  \caption{Spectrum of Cyg X-1 in the Suzaku XIS0+XIS3 (black), PIN (red)
    and GSO (green) detectors, from Miller et al (2012). The ratio to
    the best fitting model (Table 1) is shown below. The data have
    been rebinned for display purposes only.   }
\end{figure}

\begin{figure}
  \centering
  \includegraphics[width=0.7\columnwidth,angle=-90]{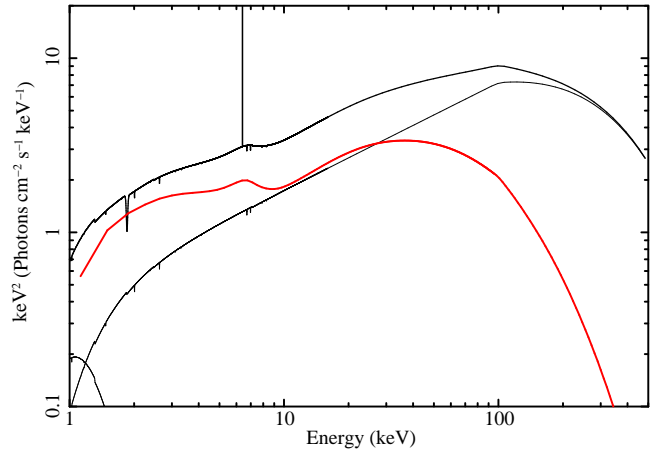}
  \caption{Best-fitting spectral model plotted as $EF_E$. The
    relativistically-blurred reflection component is shown as the red
    curve. The 100~keV kink introduced by {\sc HIGHECUT} has no
  effect on our spin results. }
\end{figure}

We show the model spectrum over the 2--65~keV band in the top panel of
Fig.~3. The inner radius has been stepped from 1.3 to 1.9, and then to
$400 r_{\rm g}$ and the innermost emissivity index $q= 9$. The effect
of fixing $q$ at 3 is shown in the lower panel: the model is now less
sensitive  to the innermost radii.

\begin{table}
  \caption{Values of  key model parameters used in the spectral fitting.}
\centering
\begin{tabular}{llll}
  	\hline
   	\textbf{Component} & \textbf{Parameter} & \textbf{Value} & 
\textbf{$q$ fixed}\\
	        \hline
	powerlaw  & Photon index, $\Gamma$ & $1.37^{+0.014}_{-0.011}$ & 
$1.39^{+0.015}_{-0.009}$        \\
	\hline
	kdblur  & Inclination, deg & $23.7^{+6.7}_{-5.4}$ & $39.8^{+3.0}_{-4.2}$ \\
	& $R_{\rm in}$\,r$_{\rm g}$ & $1.60^{+0.01}_{-0.16}$ & $<1.66$ \\
        & Index, $q_1$ & $>6.8$ & 3 \\
        & Index, $q_2$ & $2.75\pm0.15$ & 3 \\
	& $R_{\rm break}$,\,r$_{\rm g}$ & $4.0\pm1.1$ &  \\
               \hline
	reflionx  	& Iron abundance / solar &
        $1.42^{+0.64}_{-0.15}$ & $0.89^\pm0.12$ \\
	& Ionisation parameter, $\xi$ & $1765^{+245}_{-178}$ & $2960^{+202}_{-376}$ \\
        & $\ergcmps$ & & \\
  	\hline
       &  $\chi^2/{\rm dof}$ & $2217/2419$ & $2411/2421$\\
        \hline
\end{tabular}

\label{par.tab}

\end{table}

\begin{figure}
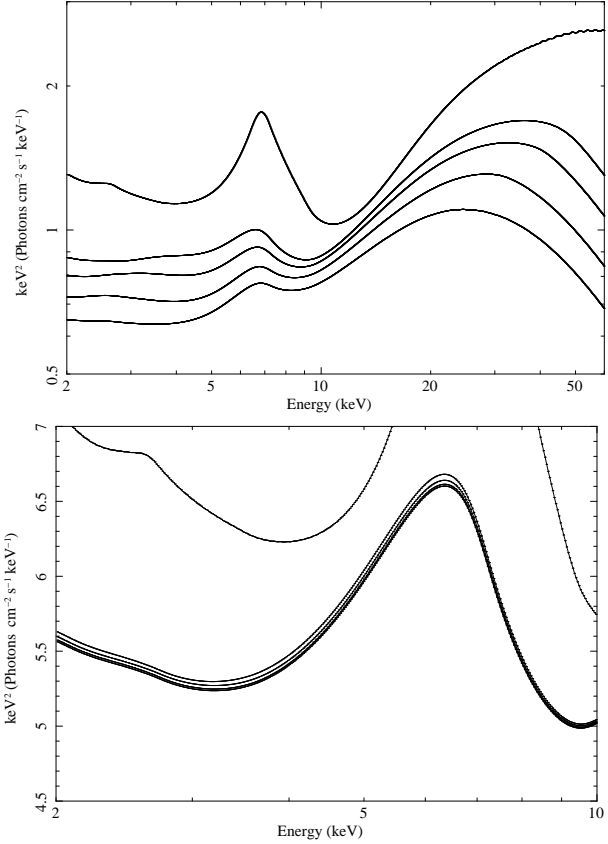

  \centering
  \includegraphics[width=0.65\columnwidth,angle=-90]{testall.ps}
  \includegraphics[width=0.67\columnwidth,angle=-90]
{newmodplt_qeq3_25_1800_gam14_all.ps}  
\caption{(Top) Model reflection spectra, relativistically blurred for
  inner radii of 1.3, 1.5, 1.7, 1.9 and 400 $r_{\rm g}$ with an
  emissivity index $q$ of 9. (Bottom) Similar to above but with $q=3$
  and expanded scales. Note the reduced sensitivity to small values of
  $r_{\rm in}$.}
\end{figure}

\begin{figure}
  \centering
  \includegraphics[width=1\columnwidth,angle=0]{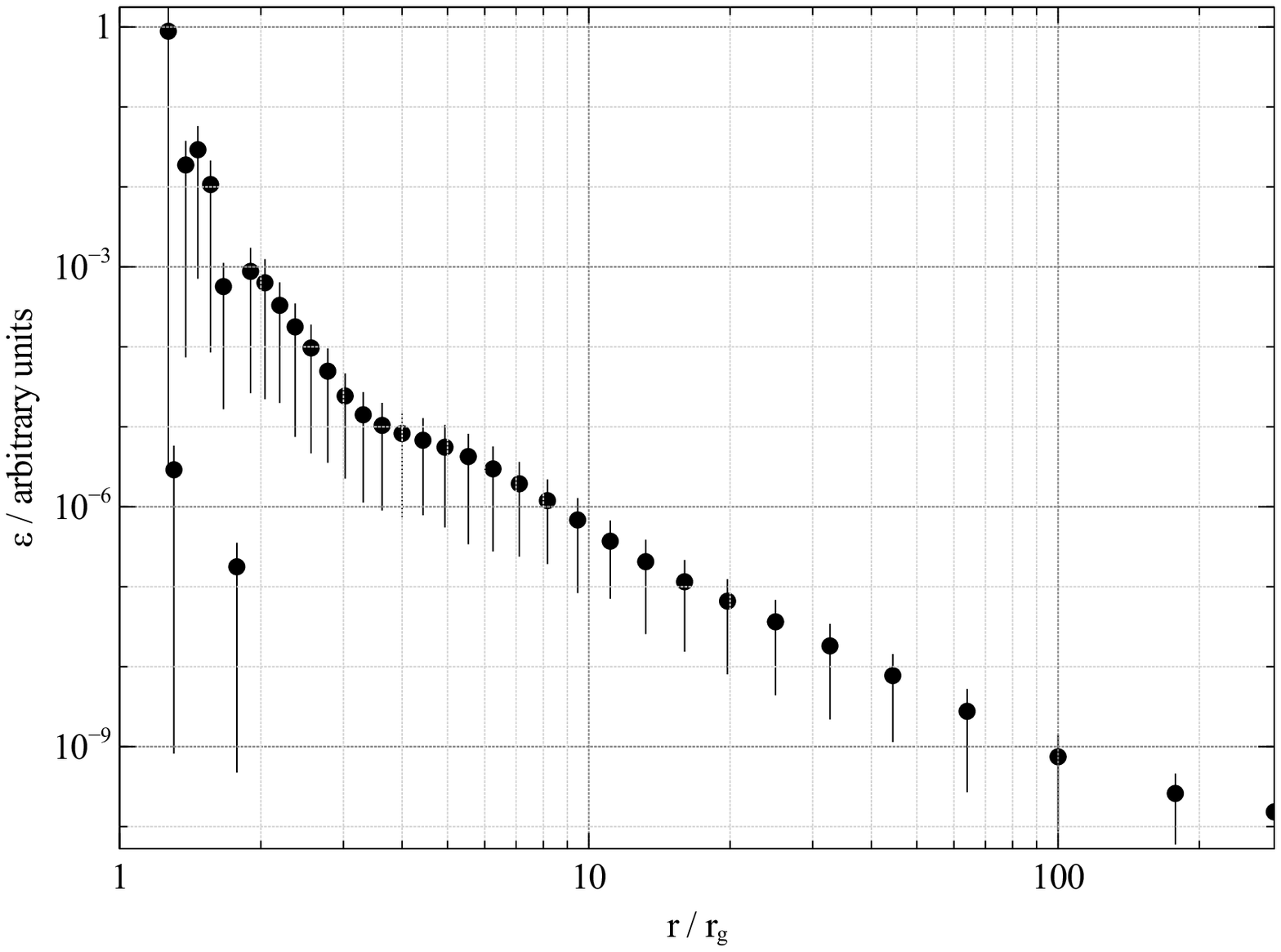}
  \caption{Emissivity profile obtained by fitting the blurred
    reflection model to small increments in radius. The profile has
    the characteristic steep -- flat-- moderate profile expected from
    irradiation by a source at a height $h=5-7r_{\rm g}$ above the
    black hole. The innermost annulus with a significant detection is
    at $r\sim 1.4r_{\rm g}$. }
\end{figure}

We next fit the spectrum with a model in which the blurring is carried
out on many small contiguous annuli, simultaneously. This is similar
to the approach taken by Wilkins \& Fabian (2011) for the AGN
1H\,0707-495. Once again, the emissivity has a steep index at small
radii before breaking to a flatter part, then dropping further close
to the Newtonian expected value of 3.  Simulations produced by ray
tracing in the strong gravity regime revealed in Wilkins \& Fabian
(2012) show that this triple power-law shape is expected from an
irradiated disc around a rapidly spinning black hole. The outer break
suggests that the height of the irradiating source above the disc is
around $5-7 r_{\rm g}$.  The  results show that emission from
within $2r_{\rm g}$ is definitely required.

The model which was used to fit the data from the XIS, PIN and GSO
simultaneously uses {\sc kdblur2f}, which is a broken power-law
capturing the essence of Fig.~4.  We investigated the dependence of
the inner radius $r_{\rm in}$ on various key parameters, the iron
abundance and the inclination (Figs.~5--7).
 

\begin{figure}
  \centering
  \includegraphics[width=0.7\columnwidth,angle=-90]{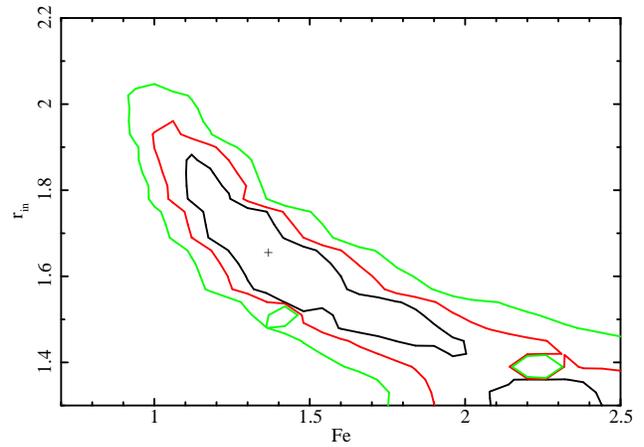}
  \caption{Effect of varying the inner radius $r_{\rm in}$ as a
    function of iron abundance $Fe$. }
\end{figure}

\begin{figure}
  \centering
  \includegraphics[width=0.7\columnwidth,angle=-90]{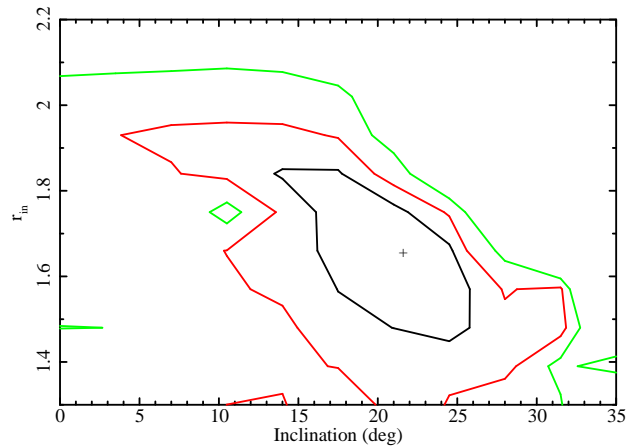}
  \caption{Effect of inclination on  inner radius $r_{\rm in}$. }
\end{figure}

The data require an iron abundance which is slightly supersolar, an
inclination less than $\sim30$~deg and an intermediate ionization parameter
around 1760. All results strongly point to an inner radius less than
$2 r_{\rm g}$, even if $q$ is fixed at 3 (Table 1, Fig.~7). 

As a test, we have fitted the PIN data alone, since this dataset is
sensitive to the reflection hump. The iron abundance is fixed at unity
(the depth of the edge in the XIS data indicates that the abundance
cannot be either very low or very high). We also fix the inclination
at 27 deg, obtaining again $r_{\rm in} < 2 r_{\rm g}$. This confirms
that fitting the reflection hump alone can in principle measure spin.

The reflection fits presented here all strongly point to the spin of
the black hole being high. Using the broken power-law emissivity model
model {\sc relconvf} (a faster modification by JSS of {\sc relconv} of
Dauser 2010, 2012) yields $a=0.97^{+0.014}_{-0.02}$ at the 90 per cent
confidence level. The results for the other parameters are similar to
those shown in Table~1.

We have also used the Novikov \& Thorne (1973; see also Page \& Thorne
1974) emissivity profile which is appropriate if the corona is
immediately above the disc. The emission then varies with radius in
the manner expected from a radiatively-efficient thin disc. The
resulting fit has $\chi^2=2422/2422$, significantly worse than the
earlier best fit. High spin ($a>0.98$) is again required and the
inclination is higher ($i\sim 39$~deg).

\subsection{The Newtonian  power-law emissivity profile, $q=3$}

Duro et al (2011) measure the broad iron line of Cygnus X-1 as seen by
XMM-Newton in the EPIC-pn modified timing mode. They report that the
black hole is spinning close to maximal with value
$a=0.88^{+0.07}_{-0.11}$, on adopting the Newtonian, single power-law
emissivity profile with index fixed at $q=3$. A fit with $q$ free gave
$q=10^{+0}_{-6}$ and $a=-0.1\pm0.4$. They rightly dismiss the
statistically more probable, high $q$ -- low spin, solution as making
no physical sense, since high $q$ is only applicable at high spin.

Fitting our spectra with $q$ fixed at 3 leads to a much
worse (although acceptable) fit with $\Delta\chi^2=+194$ (Table~1). The inner
radius is inferred to be at less than $1.66 r_{\rm g}$, the
inclination is above 35~deg (which is inconsistent with the optical
result) and the ionization parameter is higher at nearly 3000. If the
inclination is fixed at 25~deg, then $r_{\rm in}<1.9r_{\rm g}$.  The
conclusion that the black hole is rapidly spinning remains robust.

The spectral residuals obtained from using a single fixed value of $q$
and a broken power-law with variable indices are shown in Fig.~7. The
large difference in $\chi^2$ is mostly due to the red wing of the line
feature at 2.5--3.5~keV and (to a much lesser extent) the edge at
$6.5-7.5\keV$. The residuals for fixed $q$ are at about 2 per cent and
increase to 3 per cent or more if $r_{\rm in}$ is increased to $3
r_{\rm g}$ and beyond. This demonstrates where our information is
coming from and shows that from a statistical point of view we can
probe the strong gravity regime. 

\begin{figure}
  \centering
  \includegraphics[width=0.7\columnwidth,angle=-90]{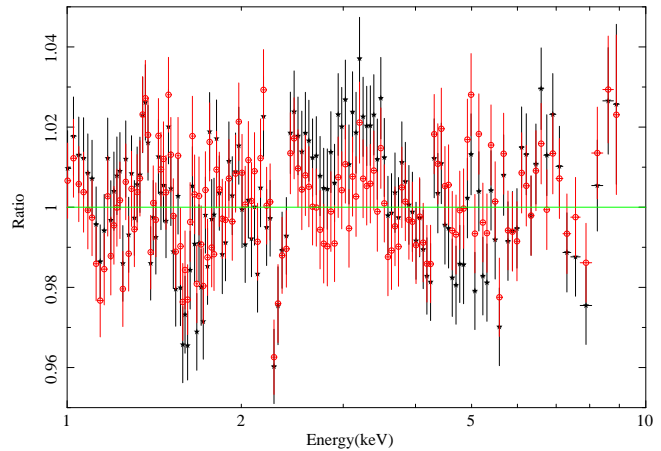}
  \caption{Ratio of (heavily binned) spectral datapoints to the best
    fitting model with a single power-law emissivity profile (index $q$
    fixed at 3: black points) and to the broken power-law model with
    variable indices (red points).  }
\end{figure}

\begin{figure}
  \centering
  \includegraphics[width=0.95\columnwidth,angle=0]{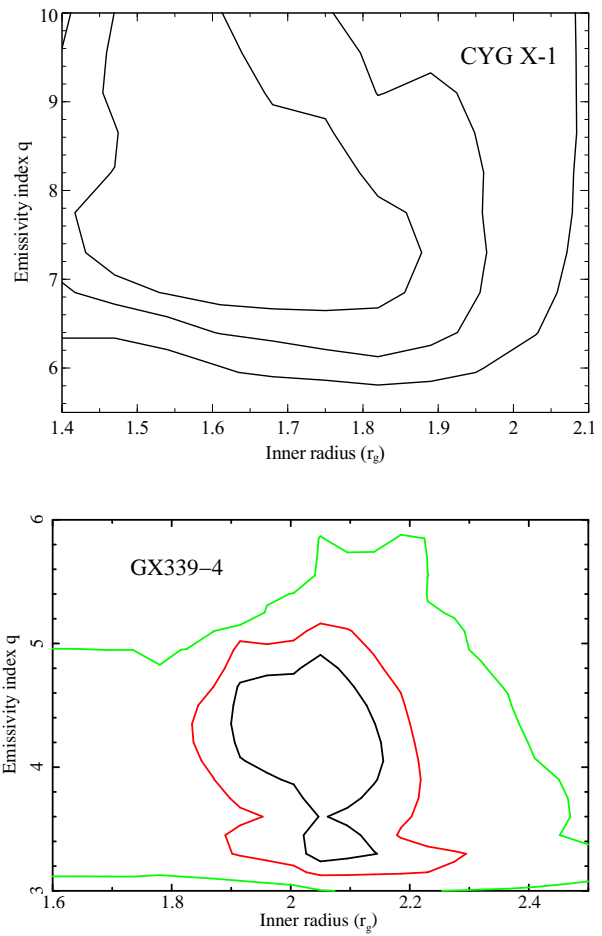}
  \includegraphics[width=0.67\columnwidth,angle=-90]{339_con_jss.ps}
  \caption{ Variation of inner emissivity index $q_1$ on inner radius
    $r_{\rm in}$ for Cyg X-1 (top) and GX339-4 (bottom). A broken
    power-law emissivity profile is used with the outer index, $q_2$,
    fixed at 3, for both plots.  }
\end{figure}

A single power-law emissivity has only limited validity. It is clear
from the fitted emissivity profile in Fig.~4 that a slope of 3 is a
fair approximation from $r=2r_{\rm g}$ outward. It underestimates the
profile within $r=2r_{\rm g}$ and is therefore a poorer probe of the
innermost region (see Fig.~3).  If $q=3$ is used, then it will likely
yield an upper limit to the inner radius and thus a lower limit on the
spin. Doubly broken power-law profiles are required when fitting for
the inner radius and/or spin when the spin is high ($r_{\rm
  in}<2r_{\rm g}$ and $a>0.94$).

Cyg X-1 is a good example where a steep inner profile is needed
(Fig.~8). In contrast, the source GX339-4, which has an inner radius
of $\sim2r_{\rm g}$ and thus moderate spin, requires only a mildly
steep index $q$ of around 4 (Fig.~7). (We use the XMM data of the
low-state observation of GX339-4 from Reis et al 2008 for this
figure.)

The higher state of Cyg X-1 during the Duro et al (2011) observations,
in which the black body component contributes to about 3~keV, may
complicate variable $q$ measurements.

\begin{figure}
  \centering
  \includegraphics[width=0.65\columnwidth,angle=-90]{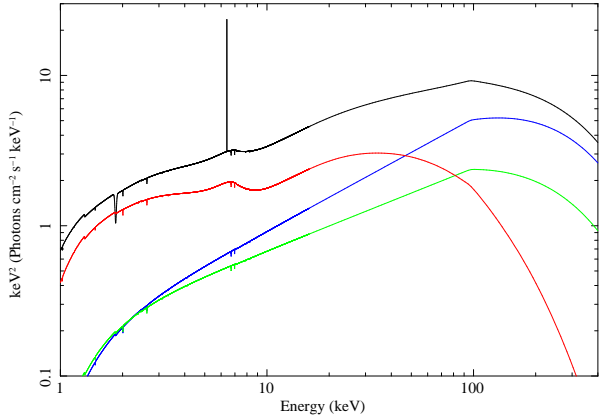}
  \caption{Model spectrum with two power-law components. The index of
    the incident spectrum for the reflection component equals that of the
    steeper power-law.}
\end{figure}

\section{Modifications to allow for the time variability of Cyg X-1}

In this Section we consider what modifications can be made to the
spectral model developed here to allow for the complex time
variability observed in Cyg X-1. The aim is not to build a complete
model which accounts for everything that has been observed but to
indicate the directions in which modifications are needed. 

Study of the rapid variability of Cyg X-1 requires high count rates
and thus instruments with large collecting areas, such as EXOSAT
(Belloni \& Hasinger 1990), Ginga (Miyamoto et al 1988, 1989;
Gierli\'nski et al 1997) and RXTE (Nowak et al 1999; Revnivtsev,
Gilfanov \& Churazov 1999; Gilfanov, Churazov \& Revnivtsev 2000;
Pottschmidt et al 2003).
The power spectrum is flat up to about 0.02~Hz, above which it drops
with frequency $f$ as $f^{-1}$ up to a few Hz where it steepens to
$f^{-2}$. Time lags are seen over the frequency range 0.1--30~Hz in
which soft bands lead hard bands, with the longest time lags being
0.05~s.  As a light crossing time, this corresponds to a distance of
$\sim 1000 r_{\rm g}$, which is much larger than the radius of any
of the regions inferred by our spectral fitting. The lag timescale
must therefore be some form of propagation time. In the low state,
such as observed here, the iron ``line'' is found to be variable up to
a frequency of a few Hz (Revnivtsev, Gilfanov \& Churazov 1999). This
is too low a frequency to be compatible with any light crossing time
effects expected in our spectral model. 

Several models have been made for the spectral variability and time
lags. Some early models are reviewed by Poutanen (2001). A model by
Lyubarskii (1997) in which fluctuations are generated by and propagate
inward through the disc has been extended by Arevalo \& Uttley (2007)
to explain many of the features observed. Most recently the hard state
lags have been modelled as propagating oscillations in a hot inner
flow in the central region of an accretion disc truncated at about $20r_{\rm
  g}$ (Kawabata \& Mineshige 2010), prompted by results of an earlier
Suzaku observation of Cyg X-1 reported by Makishima et al (2008). In
that work the corona was modelled as having two optical depths, which
allow for softer and harder continua. Reflection was dealt with by a
phenomenological model involving cold unblurred reflection and a broad
gaussian iron line.

The results obstained from the fits presented here reveal an inner
radius of less than $2r_{\rm g}$ strongly irradiated by a corona
situated at $\sim 5-7~r_{\rm g}$. This is incompatible with an
extensive inner corona. The single power-law continuum used here does
not allow for time lags unless it varies in slope on the required
timescales.  To allow for slope variations we have fitted the spectrum
with two power-law continua, the steeper of which gives rise to the
reflection. The spectral fitting is degenerate provided that the sum
of the two power-laws approximates the single power-law shown in
Fig.~2. For illustration purposes we have selected a pair of spectral
indices which give a good fit (Fig.~9; Table 2). Reflection
continues to dominate the spectrum below 10~keV.

\begin{table}
  \caption{Values of  model parameters obtained from the best fit with
    two power-law components of spectral indices $\Gamma_{\rm s}$ and
    $\Gamma_{\rm h}$. The incident spectrum for the reflection has
    index $\Gamma_{\rm s}$.}
\centering
\begin{tabular}{llll}
  	\hline
   	\textbf{Component} & \textbf{Parameter} & \textbf{Value} 
\\
	        \hline
	powerlaw  & Photon index, $\Gamma_{\rm s}$ & $1.45$f      \\
	powerlaw  & Photon index, $\Gamma_{\rm h}$ & $1.25$f    \\
	\hline
	& $R_{\rm in}$\,r$_{\rm g}$ & $1.63$  \\
               \hline
	reflionx  	& Iron abundance / solar &
        $1.28$  \\
	& Ionisation parameter, $\xi$ & $1775$  \\
        & $\ergcmps$ & & \\
  	\hline
       &  $\chi^2/{\rm dof}$ & $2219/2419$\\
        \hline
\end{tabular}

\label{par.tab}

\end{table}

A simple physical interpretation arises if the source of the steeper
power-law is quasi-static and situated close to the centre of the
disc, therefore producing the reflection.  The harder component could
originate from further out within the moving jet; both soft source
photons from the disc and reflection produced by the continuum are
reduced by beaming (Beloborodov 1999).  Time lags then occur if
variations propagate up the jet from the inner, quasistatic and softer
component to the outer, jetted harder component. They need not
propagate at jet speed, but could be due to changes in the magnetic
structure or collimation of of the jet. The power in these components
and the jet is presumed to derive from the accretion disc and be
transferred by magnetic fields coupled to a range of radii, so varying
in a manner which is consistent with the Lyubarskii model.

The lack of rapid reflection variations reported by Revnivtsev et al
1999) could in part be due to light bending effects implicit in the
proximity of the inner, softer, power-law component to the black hole
(Miniutti \& Fabian 2004), if some of the rapid variability is due to
(small) changes in height of that inner component.

Further development of such models is beyond the scope of this paper,
which is focussed on the low state X-ray spectrum of Cyg X-1.

\section{Discussion}

We have used the X-ray Reflection Fitting Method to obtain a robust
measurement of the dimensionless spin parameter of the black hole in
Cyg X-1 of $a=0.97^{+0.014}_{-0.02}$. Our result agrees with the
continuum fitting value of Gou et al (2011) and yields an even higher
spin than the reflection measurement of Duro et al (2011). Cygnus X-1
definitely appears to have high spin. 

We argue that non-Newtonian values must be used for the inner
emissivity profile of the reflected emission of the innermost disc
when the spin is high ($r_{\rm in}<2r_{\rm g}$ and $a>0.94$).

The Reflection Fitting method relies on detection of the innermost
radius of the disc which is dense enough for reflection to occur. We
assume that this is the ISCO, for within that radius the plunge orbits
mean that the gas density drops rapidly so the reflection signal
vanishes. Although magnetic fields may have some effect here, we
consider it reasonable to assume that the dense parts of the disc that
give the reflection spectrum are not seriously affected (see
discussion and simulations in Reynolds \& Fabian 2008 and Shafee et al
2008). There is nevertheless a systematic uncertainty here which
simulations need to resolve.

\begin{figure}
  \centering
  \includegraphics[width=0.8\columnwidth]{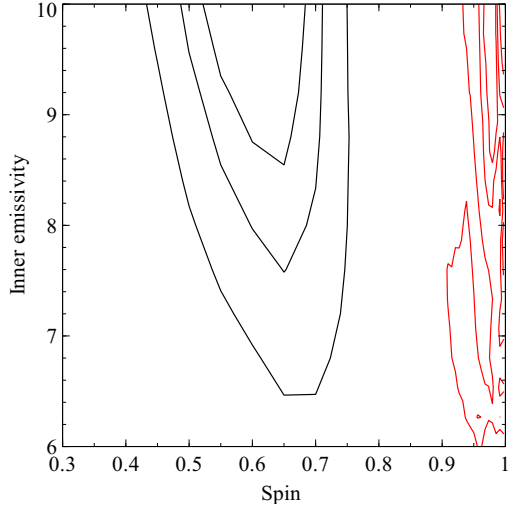}
  \caption{Emissivity -- spin contours for (left, black) a simple
    power-law profile and (right, red) a broken power-law profile with
    the outer index fixed at 3. The minimum $\chi^2$ for the fit using
    the broken power-law is about 150 below that for the fit using a
    single power-law emissivity profile.  The results in Table
    1 were obtained with the outer index free.}
\end{figure}

The reflection fraction ${\cal R}\sim 0.7$ is low compared with the
simple expectation of a source close to the black hole at
$h\sim5r_{\rm g}$. Light bending should give ${\cal R}\sim 2$
(e.g. Fabian et al 2011). The observed radio emission (Miller et al
2012) means that a jet is operating, so it is plausible that the
emission model is more complex, as discussed in Section 3. We suggest
there that irradiation of the disc by the quasi-static base of the jet
at a $5-7r_{\rm g}$ gravitational radii is responsible for the
reflected emission, whereas the observed power-law continuum is
dominated by higher, faster parts of the jet which mildly beam
emission in our direction.

Systematic uncertainties in the models and (relative) spectral
calibration can also affect the details of our fits and thus results,
particularly of the innermost region. The Continuum Fitting method is
also subject to systematic uncertainties in both models and spectral
calibration (relative and absolute).

Wide agreement on the high spin of Cygnus X-1 boosts confidence in
spin measurements from the Reflection Fitting method of other objects
as well as AGN, where it is the only method yet available.

The spectral fits strongly indicate that the innermost radius of the
disc lies within $2r_{\rm g}$. Relativistic effects are strong there
and a simple Newtonian assumption for the emissivity profile is
inappropriate. The effects are inescapable and involve both special
relativity, since the velocity of the reflecting matter on the disc is
high, and general relativity, through gravitational redshift and light
bending (Wilkins \& Fabian 2010, 2012).

Theoretical emissivity profiles due to point sources at different
heights on the rotation axis above the disc plane are shown in Fig.~12
of Fabian et al (2011). When the source height exceeds
$3r_{\rm g}$, the emissivity profile breaks at radius $r\sim h$ to the
Newtonian value of 3 outside and flatter within. The inner profile
rapidly steepens to a high value within $2r_{\rm g}$, if the disc
extends in that far.

We recommend that $q=3$ be used initially. Should this lead to high
spin being suspected then a power-law broken at least once, with a
high inner index ($q_1>3$), should be used. This is illustrated in
Fig.~10 where we have fitted our Cyg X-1 data by a single power-law
emissivity of free index. It shows the best fit being of high index
and relatively low spin ($q\sim 10,\ a\sim0.6$), which is rejected by
the physical sense argument of Duro et al (2011). If instead we use a
broken power-law emissivity fixing the outer index $q_2=3$, then we
find that the inner emissivity $q_1$ remains high and the spin rises
to be very close to maximal ($a>0.95$). This now makes physical sense.
Allowing the outer index to be free gives the results of Table~1.

\section*{Acknowledgements}
We thank the referee for comments leading to Section 3.  ACF thanks
the Royal Society for support.  RCR thanks the Michigan Society of
Fellows and NASA for support through the Einstein Fellowship Program,
grant number PF1-120087.


\end{document}